\documentclass{ws-ijmpcs}

\begin{document}

\markboth{V. Gori}
{The CMS High Level Trigger}

\catchline{}{}{}{}{}
%
%%%%%%%%%%%%%%%%%%%%%%%%%%%%%%%%%%%%%%%%%%%%%%%%%%%%%%%%%%%%%%%%%%%%

\title{The CMS High Level Trigger
%\footnote{For the title, try not to use more than
%3 lines. Typeset the title in 10 pt roman, uppercase and
%boldface.}
}

\author{Valentina Gori
%\footnote{
%valentina.gori@cern.ch
%}
}

\address{Universit\`a degli Studi \& INFN Firenze, via G. Sansone 1, Sesto Fiorentino (FI), Italy\\
%\footnote{State completely without abbreviations, the
%affiliation and mailing address, including country. Typeset in 8 pt
%italic.}\\
valentina.gori@cern.ch}

%\author{SECOND AUTHOR}

%\address{Group, Laboratory, Address\\
%City, State ZIP/Zone, Country\\
%second\_author@domain\_name}

\maketitle

\begin{history}
\received{Day Month Year}
\revised{Day Month Year}
\end{history}

\begin{abstract}
%The abstract should summarize the context, content
%and conclusions of the paper in less than 200 words. It should
%not contain any references or displayed equations. Typeset the
%abstract in 8 pt roman with baselineskip of 10 pt, making
%an indentation of 1.5 pica on the left and right margins.
The CMS experiment has been designed with a 2-level trigger system: the Level 1 Trigger, implemented on custom-designed electronics, and the High Level Trigger (HLT), a streamlined version of the CMS offline reconstruction software running on a computer farm. A software trigger system requires a tradeoff between the complexity of the algorithms running on the available computing power, the sustainable output rate, and the selection efficiency. Here we will present the performance of the main triggers used during the 2012 data taking, ranging from simpler single-object selections to more complex algorithms combining different objects, and applying analysis-level reconstruction and selection. We will discuss the optimisation of the triggers and the specific techniques to cope with the increasing LHC pile-up, reducing its impact on the physics performance.

\end{abstract}

%\ccode{PACS numbers:}

\section{The CMS trigger}

The collision rate at the Large Hadron Collider (LHC) is heavily dominated by large cross section QCD processes, which are not interesting for the physics program
of the CMS experiment. The processes we are interested in usually occur at a rate smaller than 10 Hz.
Since it is not possible to register all the events and to select them later on, because of a limited bandwith, it becomes mandatory to use a trigger system
in order to select events according to physics-driven choices.\\
The CMS experiment features a two-level trigger architecture.
The first level (L1), hardware, operates a first selection of the events to be kept, using muon chambers and calorimeter information.
The maximum output rate from L1 is about 100 kHz [\refcite{tridas}]; this upper limit is given by the CMS data acquisition electronics.
The second level, called \textit{High Level Trigger} (HLT), is implemented in software and aims to further reduce the event rate to about 800 Hz on average.
Events passing the HLT are then stored on local disk or in CMS Tier-0; about a half of these events were promptly reconstructed (within 48 hours),
while the other half have been \textit{parked} and then reconstructed along 2012.\\

\section{The High Level Trigger}
The HLT hardware consists of a single processor farm composed of commodity PCs, the event filter farm (EVF).\\
The main idea is that each HLT trigger path is a sequence of reconstruction and selection steps of increasing complexity.
The filtering process uses the full granularity data from the detector, and the selection is based on sophisticated offline-quality reconstruction algorithms.
In fact, the HLT algorithms uses a dedicated version of the software framework and the reconstruction code which is used for offline reconstruction and analysis; this online version differs from the offline one only for a different parameter configuration.\\
HLT starts from the L1 candidate, and then improves the reconstruction and filtering process by exploiting also the tracker information.
The starting selection based on the L1 information allows to reduce the rate before tracking reconstruction - a very CPU-expensive process - is performed.
In fact, the most challenging aspect is that the CMS high level trigger has to maximize the efficiency while, at the same time, keeping the CPU-time 
(not only the rate) acceptable.\\
Events are grouped into a set of non-exclusive streams according to the HLT decisions. In addition to the primary physics stream, ``stream A'', 
monitoring and calibration streams are also written. Finally, the event filter farm also collects monitoring information and makes it available to the shift crew
[\refcite{nota}].

\section{Timing and rates in 2012}

\section{HLT performance: object identification}

\subsection{Tracking and Vertexing}
\begin{figure}[htbp]
\centerline{\includegraphics[width=6cm]{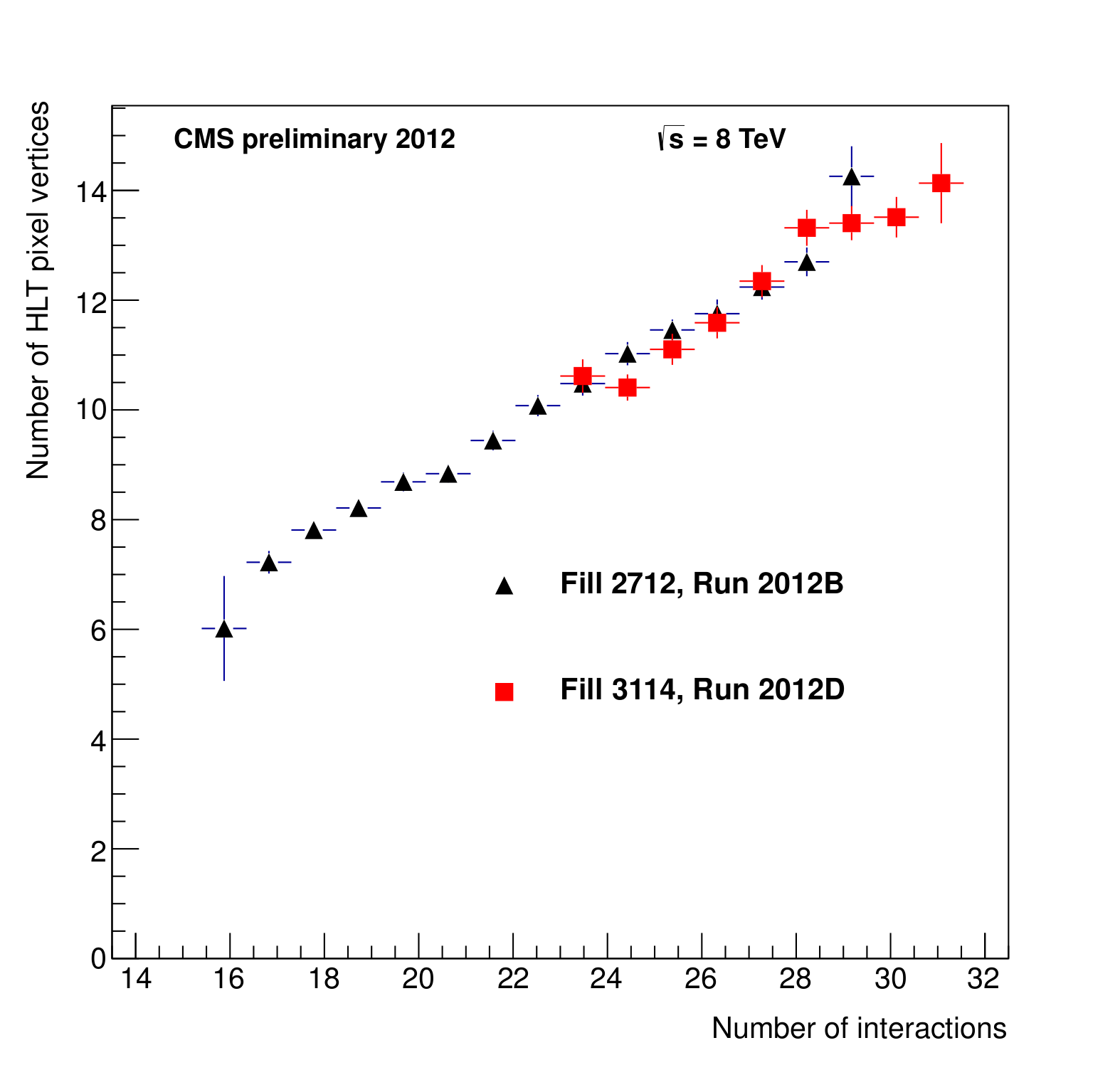}}
\vspace*{8pt}
\caption{Number of pixel vertices reconstructed at HLT.
The number of interactions is calculated from the bunch luminosity as measured by 
the forward calorimeters (HF). \label{f1}}
\end{figure}

Tracking is very important for the reconstruction at the HLT level. A robust and efficient tracking can help reconstruction 
of particles and can improve their resolution in various ways. For example, it reduces the muon trigger rate by substantially improving the momentum resolution;
energy clusters found in the electromagnetic calorimeters can be identified as electrons or photons depending on the presence of a track; 
the background rejection rate of the lepton triggers can be enhanced further by requiring that leptons 
should be isolated; it is possible to trigger on jets produced by b-quarks, by counting the numbers of tracks in a jet which have a transverse impact parameter 
incompatible with the track originating from the beam-line; it is possible to trigger on hadronic $\tau$ decays by finding a narrow, 
isolated jet using tracks in combination with the calorimeter information.\\
The pixel tracking and other track reconstruction uses about 30\% of the total HLT CPU time.
This is kept low by only performing track reconstruction when necessary (on about 4\% of total HLT events) and only after other
requirements have been satisfied to reduce the rate at which tracking must be done.\\
Pixel tracks are used to reconstruct the position of the interaction point. For vertex reconstruction, a simple gap clustering algorithm is used. 
All tracks are ordered by the $z$ coordinate of their point of closest approach to the beamspot. Then, wherever two neighbouring elements in this
ordered set had a gap between them exceeding a distance cut $z_{sep} $, this point is used to split the tracks on either side of it into separate vertices. 
In 2012 data taking, where up to 30 interactions per bunch crossing were registered, the number of reconstructed vertices still showed
a linear dependence on the number of interactions without saturating (Fig. \ref{f1}). In this Figure the real number of interactions is measured 
using the information from the forward calorimeter (HF), which covers the pseudorapidity range $3 < |\eta | < 5$.

\subsection{Muon identification}
\begin{figure}[htbp]
\centerline{\includegraphics[width=6cm]{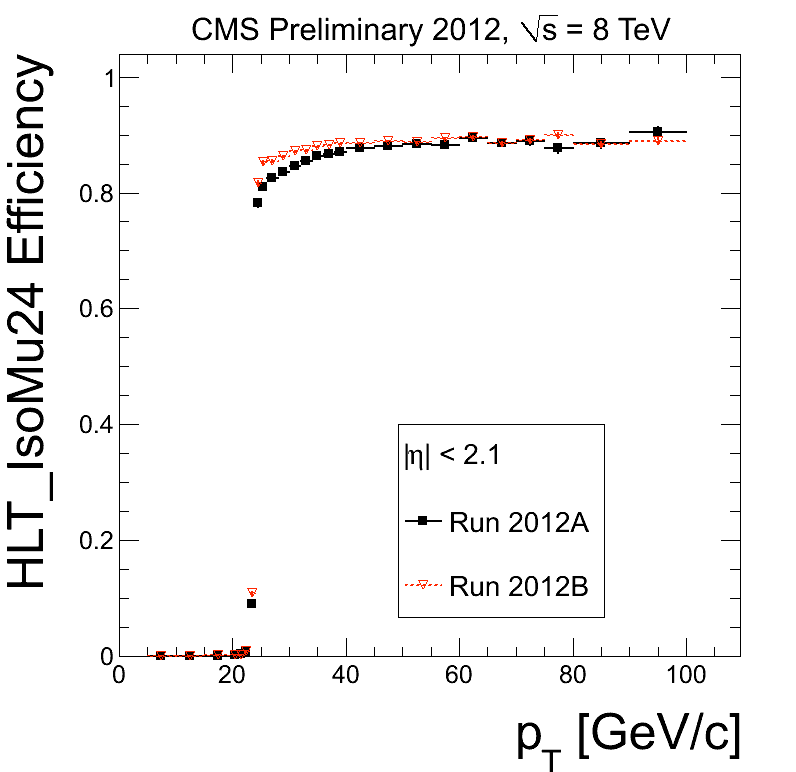}}
\vspace*{8pt}
\caption{HLT\_IsoMu24 trigger path efficiency calculated with respect to
 the offline reconstruction. \label{f2}}
\end{figure}

The muon high-level triggers at CMS combine information from the muon and the tracker
subdetectors to identify muon candidates and determine their transverse momenta, $p_T$ . The
algorithm is composed of two main steps: Level-2 (L2), which uses information from the muon
system only, and Level-3 (L3), which combines measurements from both tracker and muon
subdetectors.
In the L2, the reconstruction of a track in the muon spectrometer starts from an
initial state, called $seed$, built from patterns of DT and CSC segments. The L3
muon trigger algorithm consists of three main steps: seeding of tracker reconstruction starting
from L2 information, track reconstruction in the tracker, and combined fit in the tracker and
muon systems.
In Fig. \ref{f2} the efficiency turn-on curve for an isolated trigger path 
requiring a single muon with a $p_T$ threshold of $24$ GeV is shown. 
The isolation of L3 muons is evaluated combining information from the silicon tracker and the electromagnetic (ECAL) and hadronic (HCAL) calorimeters. 
Tracks are reconstructed in the silicon tracker in a geometrical cone of size $\Delta R = \sqrt {\Delta \eta ^2 + \Delta \phi ^2} = 0.3$ around the L3 muon. 
In the same cone, ECAL and HCAL deposits are reconstructed. To reduce the dependance of the isolation variable on the pileup of $pp$ collisions, 
the calorimeter deposits are corrected for the average energy density of the event.

\subsection{Particle Flow jets}
\begin{figure}[htbp]
\centerline{\includegraphics[width=6cm]{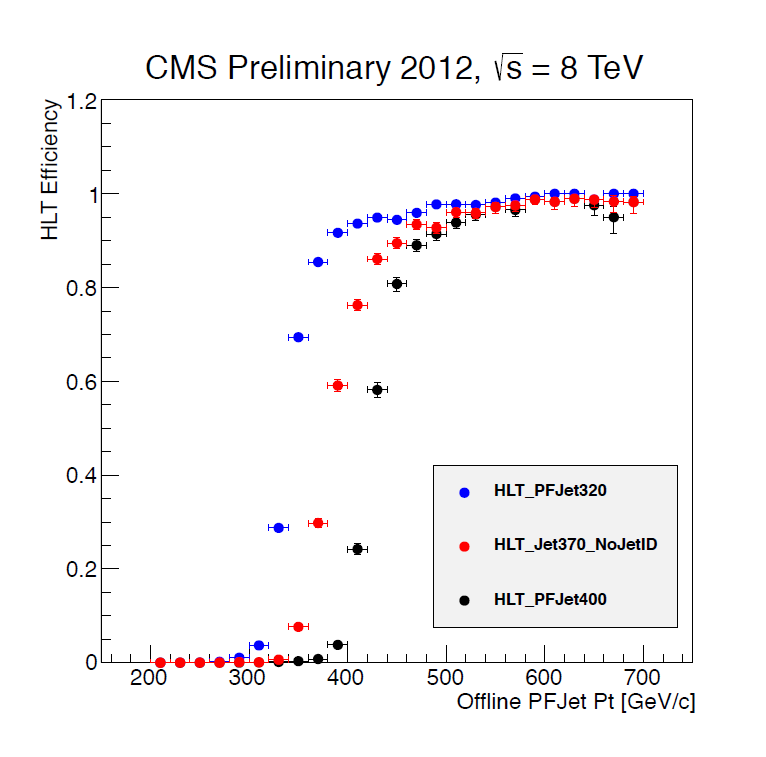}}
\vspace*{8pt}
\caption{Turn-on curve measured Vs. offline Particle Flow jet pT.
Trigger efficiency measured on an unbiased data sample from Run2012C. \label{f3}}
\end{figure}

At the HLT, jets are reconstructed using the anti-kT clustering algorithm with cone size $R =0.5$ [\refcite{antikT}]. 
The inputs for the jet algorithm can be the calorimeter towers (called CaloJet), or the reconstructed Particle Flow objects (called PFJet). 
The Particle Flow tecnique allows to use the information from all the detectors and to combine them together to reconstruct the objects [\refcite{PF}].
In 2012, most of the jet trigger paths use PFJet. Because of the significant CPU consumption of the Particle Flow algorithm at the HLT, PFJet
trigger paths have a pre-selection based on the CaloJet before the particle flow objects will be reconstructed, and PFJets will be formed. 
The matching between CaloJet and PFJet is also required in single PFJet paths.
In Fig. \ref{f3} the efficiency turn-on curve of three different trigger paths requiring PFJets with different $p_T$ thresholds are shown.

\subsection{$b$-tagging}
\begin{figure}[htbp]
\centerline{\includegraphics[width=6cm]{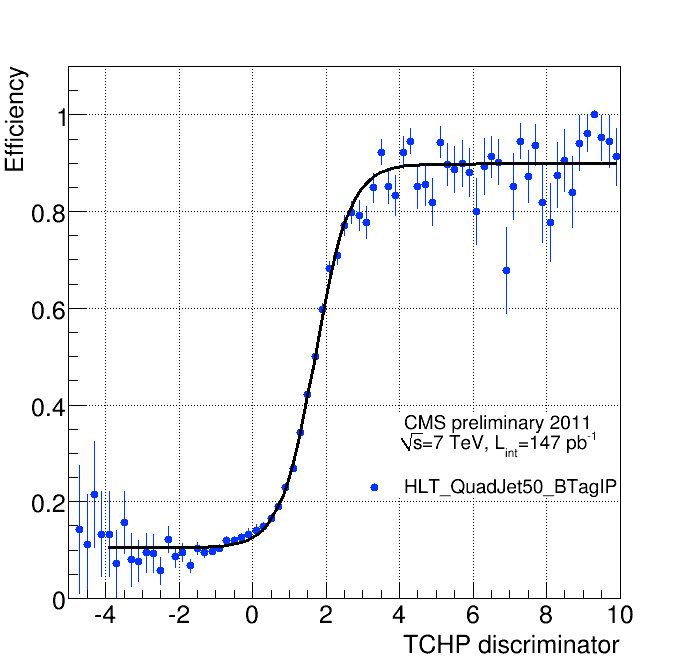}}
\vspace*{8pt}
\caption{Turn-on curve of the Track Counting High Purity (TCHP) discriminant efficiency at HLT, with respect to the same variable 
computed offline.\label{f4}}
\end{figure}

The precise identification of $b$-jets is crucial to reduce the large
backgrounds at the LHC. In CMS, using algorithms for $b$-tagging jets, this background can
already be highly suppressed at the HLT, giving lower trigger rates with large efficiency.
Algorithms for $b$-tagging exploit the fact that B hadrons typically have large decay lifetimes and
the presence of leptons in the final state compared to those from light partons and $c$ quarks. As
a consequence, tracks and vertices are largely displaced with respect to the primary vertex.
The Track Counting (TC) algorithm uses the impact parameter (IP) significance ($\sigma (IP) / IP$) of the tracks in the jets as
a discriminant to distinguish $b$-jets from other flavours. 
In Fig. \ref{f4} the turn-on curve for the Track Counting discriminant with a High Purity requirement is shown; the online cut for this path is at TCHP$=2$.
The discriminant is defined as the third highest impact parameter significance for the tracks associated to a jet.

\newpage
\section{References}
% 
% References are to be listed in the order cited in the text in Arabic
% numerals.  They should be listed according to the style shown in the
% References. Typeset references in 9 pt roman.

% References in the text can be typed in superscripts,
% e.g.: ``$\ldots$ have proven\cite{autbk}\cdash\cite{rvo} that
% this equation $\ldots$'' or after punctuation marks:
% ``$\ldots$ in the statement.\cite{rvo}'' This is
% done using LaTeX command: ``$\backslash$cite\{name\}''.
% 
% When the reference forms part of the sentence, it should not
% be typed in superscripts, e.g.: ``One can show from
% Ref.~\refcite{autbk} that $\ldots$'', ``See
% Refs.~\refcite{jpap}--\refcite{autbk}, \refcite{rvo}
% and \refcite{pro} for more details.''
% This is done using the LaTeX
% command: ``Ref.$\sim\backslash$refcite\{name\}''.

%\begin{thebibliography}{000} %for 3 digits
%\begin{thebibliography}{00}  %for 2 digits


\begin{thebibliography}{0}    %for 1 digit
% 
% %%journal paper
% \bibitem{jpap} R. Loren and D. B. Benson, {\it J. Comput.
% System Sci.} {\bf 27}, 400 (1983).
% 
% %%collaboration
% \bibitem{colla} OPAL Collab. (G. Abbiendi {\it et al}.),
% {\it Eur. J. Phys. C\/} {\bf 11}, 217 (1999).
% 
% %%normal book (authors)
% \bibitem{autbk} R. Loren and D. B. Benson, {\it Introduction to String
% Field Theory}, 2nd edn. (Springer-Verlag, New York, 1999).
% 
% %%normal book (editors)
% \bibitem{edbk} R. Loren and D. B. Benson (eds.), {\it Introduction to
% String Field Theory}, 2nd edn. (Springer-Verlag, New York, 1999).
% 
% %%review volume
% \bibitem{rvo} C. M. Wang, J. N. Reddy and K. H. Lee, New set of
% buckling parameters, in {\it Shear Deformable Beams}, ed.~T. Rex
% (Elsevier, Oxford, 2000), p.~201.
% 
% %%book in a series
% \bibitem{seri} R. Loren, J. Li and D. B. Benson, Deterministic flow-chart
% interpretations, in {\it Introduction to String Field Theory},
% Ad. Series in Math. Phys., Vol.~3 (Springer-Verlag, New York, 1999),
% p.~401.

%%proceedings
\bibitem{tridas} CMS Collaboration, \textit{CMS The TRIDAS Project, Technical Design Report Vol. 2: Data
Acquisition and High-Level Trigger}, CERN LHC 02-26, CMS TDR 6 (2002).

\bibitem{nota} CMS Collaboration, \textit{The CMS Trigger System}, CMS PAPER TRG-12-001.

\bibitem{antikT} Matteo Cacciari, Gavin P. Salam, and Gregory Soyez, \textit{The anti-kt jet
clustering algorithm}, JHEP, 04:063, 2008.

\bibitem{PF} CMS collaboration, \textit{Commissioning of the Particle-Flow reconstruction in Minimum-Bias and Jet Events from pp Collisions at 7 TeV},
CMS-PAS-PFT-10-002, 2010.

\end{thebibliography}
\end{document}